\documentclass[aps,prd,twocolumn,showpacs,amsmath,nofootinbib]{revtex4}
\usepackage[dvips]{color,graphicx}
\usepackage{amsfonts}
\usepackage{amssymb}
\usepackage{latexsym}
\usepackage{footmisc}

\newcommand{\dalm}{\kern1pt\vbox{\hrule height 0.9pt\hbox{\vrule width
0.9pt\hskip 2.5pt\vbox{\vskip 5.5pt}\hskip 3pt\vrule width 0.3pt}\hrule height
0.3pt}\kern1pt}

\newcommand{\lw}[1]{\smash{\lower2.ex\hbox{#1}}}

\usepackage{xcolor,soul}
\usepackage{bm}
\usepackage{color}

\begin{document}



\title{Fundamental forces and their dynamics }
\author{Naresh Dadhich}\email{nkd@iucaa.in}
\affiliation{Inter-University Centre for Astronomy \&
Astrophysics, (IUCAA), Post Bag 4, Pune 411 007, India}

\date{\today}

\begin{abstract}
In  this essay, we wish to propose a general principle: \it{the equation of motion or dynamics of a fundamental force should not be prescribed but instead be entirely driven by geometry of the appropriate spacetime manifold, and the equation is then obtained by employing only the geometric property without appeal to an action.} The motivation for this pronouncement comes from the fact that the equation of motion of general relativity follows from the geometry of Riemannian spacetime manifold without appeal to anything else from outside. The driving differential geometric property is the Bianchi identity satisfied by the Riemann curvature tensor. Similarly it is geometry of the principal tangent bundle of fibre spacetime manifold that may account for dynamics of the gauge vector fields. It is the classical electric force for the Abelian gauge symmetry group while the non-Abelian symmetry leads to the non-Abelian forces, the weak and the strong. We shall also reflect on a unified picture of the basic forces, and the duality correspondences it may inspire.
\end{abstract}


\maketitle

\section{Introduction}

Newton's law of gravity and Coulomb's law of electric force were motivated by the conservation of flux of force across $2$-sphere ensuring conservation of mass and electric charge enclosed inside.  By synthesizing Coulomb, Ampere and Faraday, and adding his ingenious displacement current, Maxwell wrote the famous equations of motion of electromagnetic field. This synthesis of electric and magnetic fields gave rise to existence of electromagnetic wave that propagated in vacuum with the universally constant invariant velocity, the velocity of light. It then gave rise to the realization that electric or magnetic aspect is manifestation of the same force when charge (source) is at rest or moving. The both should therefore be brought on the same footing; i.e., having the same dimension. For that to happen universally for all observers, again a universally invariant velocity would be required. Such a velocity is given by the (electromagnetic wave) velocity of light. \\

Like the synthesis of electric and magnetic fields, the invariant velocity would also bind space and time into a four dimensional spacetime manifold. That would then accord to a new mechanics, special relativity. We can also envision that special relativity is driven by universalization of the Newtonian mechanics, which described motion of all massive particles, by including zero mass particles. The invariant velocity of zero mass particle is synthesized in the spacetime structure. It serves as an excellent spacetime background for the Maxwell's electrodynamics as well as for the rest of physics except, of course, gravity.  \\

Like Newtonian mechanics, how about universalizing the Newtonian gravity by including massless particles as well. That means gravity should also link to zero mass particles; i.e., it is universal with linkage to both massive as well as massless particles. It links to everything that physically exists. Note that motion of zero mass particle is a property of the spacetime and hence its interaction with any force has to be negotiated through spacetime itself. For inclusion of zero mass particles in gravitational interaction, space(spacetime) has to be curved \cite{D2010, D2011, D2016}. For universalization of gravitational force, the framework has to be enlarged from flat to curved spacetime. Gravity is thus described by spacetime curvature and hence it becomes the property of spacetime geometry. That is, the gravitational equation of motion cannot be prescribed instead it has to be determined by geometry of the spacetime. That is what happens in the Einstein gravity, general relativity (GR). In particular Newton's inverse square law is mandated by the Riemannian geometry of spacetime.  \\

It then motivates us to ask, should this not be the case for the  other fundamental forces as well ? Of course a particular force would accord to a geometry of appropriate spacetime manifold. For a universal force of gravity, spacetime manifold should be universally accessible, and hence it is the spacetime itself which is accessible and shared by all equally without any qualification. The gravitational dynamics follows from the differential geometric property -- the Bianchi identity satisfied by the Riemann curvature tensor. In GR gravitational dynamics resides in the spacetime curvature. \\

What are the other possible choices for spacetime manifold. One obvious candidate is the fibre bundle manifold of the principal tangent bundle. With the qualification of fibre bundle, this manifold is accessible only to particles having a particular charge. Again it is the curvature of the principal tangent bundle that leads through the Bianchi identity to an equation of motion of the force -- the Maxwell equation of electrodynamics. The dynamics of electric force is thus governed by the geometry of the principal tangent bundle spacetime. More generally dynamics of a gauge vector field could be described by the tangent bundle spacetime. \\

A vector field could as well be decorated by internal space indices indicating $SU(N)$ symmetry group, where $N=1$, is the Abelian symmetry of the familiar Maxwell field while $N>1$ describes non-Abelian symmetry of the Yang-Mills fields. In particular $SU(2)$ and $SU(3)$ respectively correspond to the weak and the strong forces. \\

It then turns out that the equations of motion of all the four fundamental forces follow from the geometry of the corresponding spacetime manifold without appeal to anything else from outside. The equation is obtained by using only the differential geometric property of Bianchi identity without appeal to an action. We then propose that this should be the defining property of a fundamental force. \\

In general one derives equation of motion of any force, fundamental or otherwise, by writing an appropriate action Lagrangian, which on variation with respect to the corresponding (potential) variable yields the equation of motion. This is the general principle applicable universally for all forces. For the fundamental forces it turns out that Lagrangian is written in terms of the geometric entity, curvature of specific spacetime manifold. Its variation relative to the corresponding geometric entity (metric or connection) yields the corresponding equation of motion for the force. The basic variable is then a geometric entity, curvature which satisfies the geometric property, the Bianchi differential identity. The question we wish to pose is: What does then the Bianchi identity imply for dynamics of the fundamental force ? It turns out that it determines it -- the equation of motion. This is the main purpose and motivation of this discourse. We would show that the equation of motion for all the four fundamental forces could alternatively be obtained in a much simpler and direct manner by purely employing the geometric property -- the Bianchi identity. This is because it is the curvature, which is a geometric entity, that drives the dynamics, why shouldn't it be entirely determined by the geometric property of the curvature? \\

 The paper is organised as follows: In the next section we would  derive the equations of motion of the four fundamental forces from the geometry of relevant spacetime manifold. We shall begin with the derivation of the Einstein's equation of general relativity and then take up derivation of the equations of motion for Abelian and non-Abelian vector gauge forces. Before we conclude with a discussion, we would also indulge in presenting a unified scheme of all the four forces. \\

 \section{Equation of motion of fundamental forces}

 We shall begin by the geometric characterization of homogeneous and inhomogeneous spacetime, and then show that the Einstein's gravitational equation follows naturally when spacetime is inhomogeneous.

 \subsection{Universal force -- Einstein's gravity}

 Since we wish to derive the equation of motion of a universal force, the corresponding spacetime should also be accessible to all particles without any exception. It is therefore the four dimensional spacetime manifold in which space and time are bound together by the universally invariant, the constant velocity of light. As a matter of fact, homogeneity of space and time mandates existence of the universally invariant velocity \cite{D2010, D2011,D2016}. \\

In the absence of all forces, spacetime is naturally homogeneous, and hence its geometry should also be homogeneous; i.e., Riemann curvature tensor should be homogeneous. It should be covariantly constant which means it should be written in terms of the metric, $g_{ab}$ that has vanishing covariant derivative. We thus write

\begin{equation}
R_{abcd} = \Lambda (g_{ac}g_{bd} - g_{ad}g_{bc})\, .
\end{equation}

This is a spacetime of constant curvature $\Lambda$ which could, as determined by the experiment and observation, be positive, dS or negative, AdS or zero, flat Minkowski. It is the most general characterization of "free" homogeneous spacetime. It imbibes no dynamics of any force. Dynamics of universal force emerges onlly when spacetime becomes inhomogeneous. \\

Let us appeal to the famous theorem of differential geometry,  $D^2=0$ identically where $D$ is a properly defined covariant derivative. It is known as the Bianchi identity. John Wheeler elegantly paraphrased it as {\it boundary of boundary is zero -- a conservation law}. Its familiar well known manifestations are curl of gradient and divergence of curl being zero. These are for scalar and vector fields, and their higher order analogue demands that the Bianchi derivative of the Riemann curvature vanishes identically; i.e.,
\begin{equation}
\triangledown_{[e} R_{ab]cd} = 0 \, .
\end{equation}
This is the Bianchi differential identity satisfied by the Riemann curvature tensor.\\

On taking its trace, multiplying by $g^{ac}g^{bd}$, we get
\begin{equation}
\triangledown_b G_a{}^b  = 0 \, ,
\end{equation}
where
\begin{equation}
G_{ab} = R_{ab} - \frac{1}{2} R g_{ab} \, , \, R_{ac} = g^{bd}R_{abcd}\, , \, R = g^{ab}R_{ab}\, .
\end{equation}
We have derived here a divergence free second rank symmetric tensor, known as the Einstein tensor, from the Riemann curvature which involves second derivative of the metric tensor. \\

From the above divergence equation, we can infer and write
\begin{equation}
G_{ab} + \Lambda g_{ab} = \kappa T_{ab} \, \, , \, \triangledown_b T_a{}^b = 0\,.
\end{equation}
The Einstein tensor is the creature of the Riemann curvature and hence it contains second derivative of the metric, a second order differential operator like $\triangledown^2 \Phi$, while $\Lambda g_{ab}$ is a constant of integration relative to covariant derivative. Thus far we have come purely driven by spacetime geometry with no reference or appeal to anything from outside. \\

The above equation would become the equation of motion of a force if we identify the source with the second rank symmetric tensor $T_{ab}$ with vanishing divergence. This is what would inject physical content in this otherwise purely geometric statement.  This source should make spacetime inhomogeneous for all particles without any qualification or exception, it should therefore represent a universal physical property which is shared by everything that physically exists. The obvious answer is energy-momentum. With this identification, the above equation is the Einstein's equation of gravitation -- general relativity \cite{D2010, D2011, D2016}. \\

It is remarkable that the Einstein gravity springs up naturally without any prescription from outside when spacetime is inhomogeneous. As a matter of fact we were simply exploring what is it that makes spacetime inhomogeneous without reference to gravity at all. The answer turns out that it is the presence of energy-momentum distribution that makes spacetime universally inhomogeneous and the universal force it gives rise to is uniquely the Einstein gravity in the four dimensions. \\

We have thus derived the dynamics of Einstein's gravitational force from the geometry of spacetime and employing only the geometric Bianchi identity. In particular it should be emphasized that the inverse square law is now dictated by the spacetime geometry. \\

This sets the setting for similar exploration of the other fundamental forces. \\

\subsection{Abelian Maxwell force}

Now let us explore the fibre bundle geometry of the principal tangent bundle. Its curvature is a $2$-form $F_{ab}$ which satisfies the Bianchi identity,
\begin{equation}
\triangledown_{b} ^{*}F^{ab} =0\, ,
\end{equation}
where $^{*}F^{ab}$ is the Hodge dual of $F^{ab}$. This implies that $F_{ab}$ is a curl of a vector; i.e.,
\begin{equation}
F_{ab} = \triangledown_{[a}A_{b]} = -F_{ba}\, .
\end{equation}
Here $A_{a}$ is the connection on the tangent bundle spacetime. \\

On account of the antisymmetry of $F_{ab}$, we write
\begin{equation}
\triangledown_{b}\triangledown_{a}F^{ab} = 0\, ,
\end{equation}
from which as before we write
\begin{equation}
\triangledown_{a}F^{ab} = -J^{b}\, ,\, \, \triangledown_{a}J^{a}=0\,.
\end{equation}
All this was purely geometry driven, now we inject physics into it by identifying $J^{a}$ with the conserved $4$-current, then it is  the Maxwell's equation of the electric force where $A_{a}$ is the $4$-vector potential. The other part of the Maxwell equation is however the Bianchi identity itself. \\

Like the curvature of spacetime leads to the Einstein's gravitational equation, similarly curvature of the principal tangent bundle spacetime to the Maxwell's equation of electrodynamics. Electric force has the Abelian $U(1)$ symmetry group and is therefore long range, and has massless free propagation.\\

The dynamics of the two classical long range forces, gravity and electromagnetic, follows respectively from the corresponding geometry of Riemannian spacetime and the principal tangent bundle spacetime. There is no reference or appeal to anything else from outside, and it all flows naturally from the geometry of the corresponding spacetime manifold by employing the geometric property alone. \\

\subsection{Non-Abelian Yang-Mills forces}

Non-Abelian character refers to the internal space symmetry group, $SU(N)$ for $N>1$. In $N=1$, it includes the Abelian case, $SU(1)=U(1)$. Let $\psi_{\sf a}$ denote generically the fields which transform under certain representation of the above symmetry group with generators $t_A$, represented by matrices $(t_A)_{\sf a}^{\phantom{a} \sf b}$ (the sans-serif font is used here to distinguish these from spacetime indices) \cite{weinberg}. The fields transform as $\delta \psi_{\sf a} = i \epsilon^A  (t_A)_{\sf a}^{\phantom{a} \sf b} \psi_{\sf b}$. The gauge covariant derivative in this case is introduced via the gauge potential $G^A_a$ which transforms as $\delta G^A_i = \partial_i \epsilon^A + i \epsilon^C ({\tilde t}_C)^A_{\phantom{A} B} G^B_i$ where ${\tilde t}_A$ indicates the adjoint representation. The gauge covariant derivative is then given by $(D_i \psi)_{\sf b} = \partial_i \psi_{\sf b} - i G^B_i (t_B)_{\sf b}^{\phantom{\sf b} \sf c} \psi_{\sf c}$. For such a theory, one may show that $\left( [ D_i, D_j ] \psi \right)_{\sf a} = - i (t_A)_{\sf a}^{\phantom{\sf a} \sf b} F^A_{ij} \psi_{\sf b}$ with
\begin{eqnarray}
F^A_{ij} = \partial_i G^A_j - \partial_j G^A_i + i ({\tilde t}_C)^A_{\phantom{C} B} G^B_i G^C_j
\end{eqnarray}
The key departure from the Abelian case is the last term, which makes it subtle to define a conserved charge in terms of a gauge covariant current density. Nevertheless, if one considers the gauge invariant lagrangian of the form:
\begin{eqnarray}
L = \frac{1}{4} F^A_{ij} F^{B \, ij} \delta_{AB} + L_{\rm matter}(\psi_{\sf a}, (D_i \psi)_{\sf a}, \ldots )
\end{eqnarray}
and define the matter current density by
\begin{eqnarray}
J^i_A = -i \frac{\partial L_{\rm matter}}{\partial (D_i \psi)_{\sf a}} (t_A)_{\sf a}^{\phantom{\sf a} \sf b} \psi_{\sf b} \, \, ,
\end{eqnarray}
then the equation of motion can be written as
\begin{eqnarray}
D_k F^{ki}_A = - J^i_A\, .
\end{eqnarray}
Using the transformation law for $F^{ki}_A$, it can be shown that
\begin{eqnarray}
D_i J^i_A=0\, .
\end{eqnarray}
Note that, while the above current is defined in a gauge covariant manner and satisfies a gauge covariant conservation law, it does not yield a local conserved charge.


The critical step in getting at the equation of motion is the vanishing of double divergence of the curvature $2$-form. We verify that that continues to hold good for the gauge invariant derivative as well.

Then as before we can write the equation of motion for the non-Abelian Yang-Mills force. It is the weak force for $SU(2)$ and the strong for $SU(3)$. The geometry of principal tangent bundle spacetime wraps in all gauge vector forces Abelian as well as non-Abelian.

\section{Unified view of the fundamental forces}

We shall now attempt to envisage a unified picture \cite{unified} of the four basic forces. A force is characterized by the two properties: (a) its linkage, to what does it link to, and (b) its range, how far is its reach ? \\

Let us begin with a universal force which links to everything that physically exists, massive as well as massless, and it reaches out everywhere, and hence is long range. Since it links to massless particles, which can feel it only if force curves space, rather spacetime as space and time are already bound together. It has  therefore to be described by the curvature of spacetime and its dynamics following from the Bianchi identity as we have seen above. \\

The universal force is then uniquely the Einstein's gravity, GR. \\

It is envisaged that the other three forces emerge as these two properties are peeled off one by one. If we relax the property (a) that linkage not universal but to a particular charge and retain the long range property (b). Charge has to be bipolar and hence the force would be a vector field. Since it is long range, propagator would be massless. It would then be a vector gauge field with the dynamics given by the Maxwell's equations. \\

The long range force linking to a charge is therefore uniquely the Maxwell's electric force. Einstein gravity and Maxwell's electric force are the only two long range forces. \\

Now if we relax the long range property (b). The short range would require either propagator is massive which would then link only to massive particles or the coupling is running such that it tends to zero as $r\to0$, the asymptotic freedom. The short range is achieved through the internal space structure $SU(N)$ superposed on the vector field. Then $N=2, 3$ correspond respectively to the weak and the strong force.\\

The long range forces are the classical forces and they are uniquely gravity and electromagnetic. In this envision, there can exist no other long range force. On the other hand short range ones are quantum forces where the weak  and the strong forces are identified with $N=2, 3$ for $SU(N)$ internal symmetry group. Here however the question remains open for higher $N$, keeping the possibility open for a new force. \\

It is remarkably insightful that all the four fundamental forces accord to such a simple unified picture based on the two characteristic properties, linkage and range. Further it also makes some very interesting suggestions for the duality relations. Electric and weak forces are complementary to each other, the former is long range but not universal while the latter is short range but universal in the sense of linking to all massive particles. This is a pointer to the electro-weak unification. On the other hand, gravity is universal having universal linkage as well as long range while the strong force is neither, which points to their complementarity. The most remarkable result in this line is the famous AdS/CFT correspondence \cite{AdS-CFT} which has been one of the drivers of high energy research for over two decades now. Also note that in the string theoretic understanding of the strong force, there is involvement of spin-2 massless rather than spin-1 particle.\\

These are the strong pointers for seeking duality relations between electric and weak, and gravity and strong in appropriate framework. \\

\section{Discussion}

We had set out to derive the equations of motion of all the four fundamental forces from geometry of the relevant spacetime by employing the geometric Bianchi identity. The force which is universal has to have universal linkage as well as it should reach everywhere, and hence long range. It has therefore to be governed by the geometry of spacetime manifold which is accessible to all particles; i.e., the spacetime itself. That universal force is the Einstein's gravity -- GR having its dynamics entirely governed by the Riemann curvature through the Bianchi identity. The Einstein's gravitational equation then follows without reference to anything else except of course the identification of the source on the right. \\

There is a very satisfying general feature that accord to the classical mechanics, that when spacetime is free of all forces including gravity, it is homogeneous having constant curvature. When it is inhomogeneous, Einstein's gravity emerges naturally. The constant curvature of spacetime is maximally symmetric admitting all ten Killing symmetries in the four dimensions, and it characterizes the  force free state. This means that the Lie derivative of the metric tensor along all ten Killing directions vanishes. The point to be noted is that maximal symmetry does not imply vanishing of Riemann curvature instead it simply makes it homogeneous with vanishing covariant derivative. \\

This is in contrast to the conventional textbook perspective where absence of gravity is characterized by vanishing of the Riemann curvature -- flat Minkowski spacetime. In this picture the maximally symmetric homogeneous spacetime paradoxically harbours gravitational dynamics! A homogeneous Riemann curvature cannot lead to any equation of motion because the Bianchi identity is trivial in that case. On the other hand if one identifies presence of gravity by the tidal acceleration experienced by two parallelely  propagating geodesics which would be though constant but non-zero. What would be the source of this dynamics, it could only be a constant scalar field with no dynamics.\\

On the other hand we have argued elsewhere \cite{D2016, D2022c} that the velocity of light and $\Lambda$ arise as the constants of spacetime structure as the properties of homogeneous spacetime. Then gravitational dynamics arises only when spacetime becomes inhomogeneous. Not the tidal acceleration but its inhomogeneity is indicative of gravitational dynamics. That is what we would like to surmise that gravitational dynamics is described not by homogeneous but inhomogeneous curvature. The constant curvature is therefore characterization of homogeneous spacetime providing the force free background. \\

Another question that arises is that there are also higher derivative gravitational theories and could the same procedure be employed for obtaining their equation of motion? The answer is yes, so long as the equation is of second order. This requirement uniquely singles out the Lovelock theory. Its action Lagrangian is a  homogeneous polynomial in the Riemann curvature, and yet remarkably it yields the second order equation. It is possible to define the Lovelock analogue of the Riemann tensor (which is a homogeneous polynomial in Riemann)  \cite{D2008, Kastor2012, CD2015}, and the Lovelock equation of motion then  follows by requiring that trace of the Bianchi derivative of the Lovelock Riemann tensor vanishes\footnote{Note that the Bianchi derivative of the Lovelock Riemann will not vanish and that happens only for Riemann. However for obtaining the Lovelock analogue of the Einstein tensor, vanishing of trace of the Bianchi derivative is good enough.}. Further it has also been argued that the pure Lovelock equation is perhaps the appropriate gravitational equation in higher dimensions \cite{D2012,DGJ2012}.  \\

When force links to a particular charge, it has to be bipolar so as to give total charge zero \footnote{One may ask how does total charge become zero for gravity where its charge, energy-momentum is unipolar and positive? The only way it could be neutralized is that gravitational field it produces has negative energy, which when summed over all space exactly cancels out positive charge \cite{adm}. This is why gravity is always attractive.} when summed over all charges in the Universe. It is then a vector field and its dynamics would be described by the curvature of the principal tangent bundle spacetime. \\

The absence of force would then require the principal tangent bundle spacetime to be maximally symmetric. That would mean Lie derivative of the connection on the bundle -- vector potential should vanish relative to all ten Killing vectors. It could be easily verified that that would require the curvature $2$-form to vanish; i.e., $F_{ab}=0$. That is, the tangent bundle spacetime is homogeneous only when the curvature vanishes. This is in contrast from the gravity case where it required  spacetime to be of constant (and not necessarily of zero) curvature. \\

When force is long range, it is Abelian gauge vector field with $U(1)$ symmetry. A vector field could also have internal symmetry group $SU(N)$ which essentially amounts to putting some "hooks" to  make it short range, either by making the propagator massive (the weak force) or the coupling running -- tending to zero as $r\to 0$, the asymptotic freedom (the strong force). The dynamics of vector field whether Abelian or non-Abelian follows from the curvature of the principal tangent bundle spacetime with appropriate identification of the conserved $4$-current. \\

In the unified picture of all the four forces we alluded above,  there could exist only two classical long range forces, Einstein's gravity and Maxwell's electromagnetic field, and the both have been uniquely identified. Thus there is no room for any other long range force. Any new force, if it exists, has to be short range. On the other hand the weak and strong forces are identified but there is no uniqueness about them. Any $SU(N)$ force would be very well accommodated in this overall perception. The question of new force therefore remains open in the short range regime. \\

This visualization strongly suggests duality between the electric (linkage to specific charge and long range) and the weak (universal linkage to all massive particles and short range) as they are complementary, and electro-weak unification is perhaps indicative of that. Similarly there is complementarity between the gravity, which has universal linkage as well as long range, and the strong, which is neither. The celebrated AdS/CFT correspondence \cite{AdS-CFT} is perhaps indicative of that. \\

Further it also make some very important predictions. Since the weak force links to all massive particles, the neutrino must have non zero mass and for the same reason the right handed electron must participate in the weak interaction. The latter has not yet been observed, it is a clear cut prediction that follows from this very simple intuitive picture. Also any new force, if it exists, has to be short range only. \\

Finally we would like to reiterate the main concern and proposition of this discourse as follows: \textit{Since dynamics of a fundamental force is governed by geometry of the corresponding spacetime manifold, its equation of motion also follows purely from the differential geometric property.}

It is a pleasure to dedicate this short discourse to the fond memory of my long time friend and esteemed colleague, Paddy, with much warmth and fondness. He was always interested in the fundamental questions and never hesitated in looking at things in a different and unconventional ways, usually exposing new insight and revealing vista. This is a modest offering in that vein. \\

\section{Acknowledgments}

I wish to warmly thank Abhay Ashtekar for the insightful critique of the viewpoint presented here, and to Kinjlak Lochan and Dawood Kothawala for some interesting and clarifying discussions, and the latter also for help in manoeuvering through the non-Abelian Yang-Mills derivation in Sec II C. I also wish to acknowledge the support of the CAS President's International Fellowship Initiative grant No. 2020VMA0014, and a summer visit at the Albert Einstein Institute, Potsdam-Golm.

\end{document}